\documentclass{ws-procs975x65}
\def\beq{\begin{equation}}
\def\eeq{\end{equation}}
\def\figsubcap#1{\par\noindent\centering\footnotesize(#1)}

\begin{document}

\title{Post-Newtonian cosmological models}
\author{Viraj A A Sanghai$^*$} 

\address{ School of Physics and  Astronomy, \\ Queen Mary University of London, \\ Mile End Road, London, E1 4NS, UK\\
$^*$E-mail: v.a.a.sanghai@qmul.ac.uk}

\begin{abstract}
We construct a framework to probe the effect of non-linear structure formation on the large-scale expansion of the universe. We take a bottom-up approach to cosmological modelling by splitting our universe into cells. The matter content within each cell is described by the post-Newtonian formalism. We assume that most of the cell is in the vicinity of weak gravitational fields, so that it can be described using a perturbed Minkowski metric. Our cells are patched together using the Israel junction conditions. We impose reflection symmetry across the boundary of these cells. This allows us to calculate the equation of motion for the boundary of the cell and, hence, the expansion rate of the universe. At Newtonian order, we recover the standard Friedmann-like equations. At post-Newtonian orders, we obtain a correction to the large-scale expansion of the universe. Our framework does not depend on the process of averaging in cosmology. As an example, we use this framework to investigate the cosmological evolution of a large number of regularly arranged point-like masses. At Newtonian order, the Friedmann-like equations take the form of dust and spatial curvature. At post-Newtonian orders, we get corrections to the dust term and we get an additional term that takes the same form as radiation. The radiation-like term is a result of the non-linearity of Einstein's equations, and is due to the inhomogeneity present in our model. 
\end{abstract}


\bodymatter
\section{Introduction}
The standard approach in cosmology is to assume that we can describe the large-scale expansion of the Universe using a single homogeneous and isotropic solution to Einstein's field equations. This is commonly known as the Friedmann-Lema\^{i}tre-Robertson-Walker (FLRW) solution. However, general relativity is known to be valid locally and there is no unique way to average tensors. Hence, the FLRW solution might not be the best way to approximate the large-scale expansion of the universe. This is sometimes referred to as the ``back-reaction'' problem in cosmology\cite{br}. Also, there is no perturbative scheme that works consistently on all scales, in the presence of non-linear structure. Cosmological perturbation theory works well on large scales but breaks down on small scales in non-linear regimes\cite{clarkson, rasanen}. Conversely, post-Newtonian perturbation schemes work well on small scales but break down on the very largest scales, and in the presence of strong gravitational fields \cite{Will, Ch1}.  We sidestep these issues by constructing a bottom-up approach to cosmology using the post-Newtonian approximation to gravity, the details of which can be found in Sanghai \emph{et al.}\cite{vaas}. This allows us to evaluate the effect of non-linear structure on the large-scale expansion of the Universe. Such an approach may be useful for interpreting data from future large-scale surveys such as \emph{Euclid}\cite{euclid} and \emph{SKA}\cite{ska} (Square Kilometre Array).

\section{Building A Post-Newtonian Cosmology}
\begin{figure}[h]
\begin{center}
 \parbox{2in}{\includegraphics[width=1.9in]{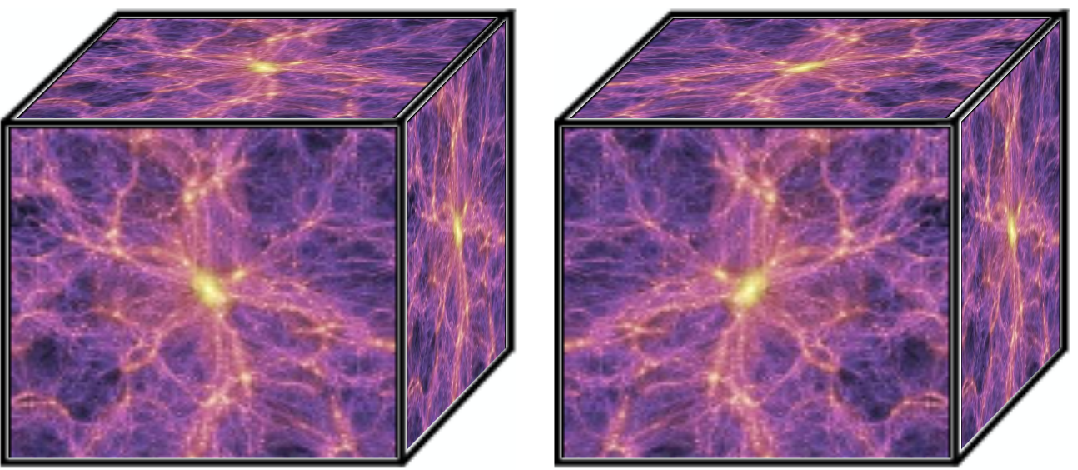}\figsubcap{a}}
 \hspace*{2pt}
 \parbox{2.1in}{\includegraphics[width=2.5in]{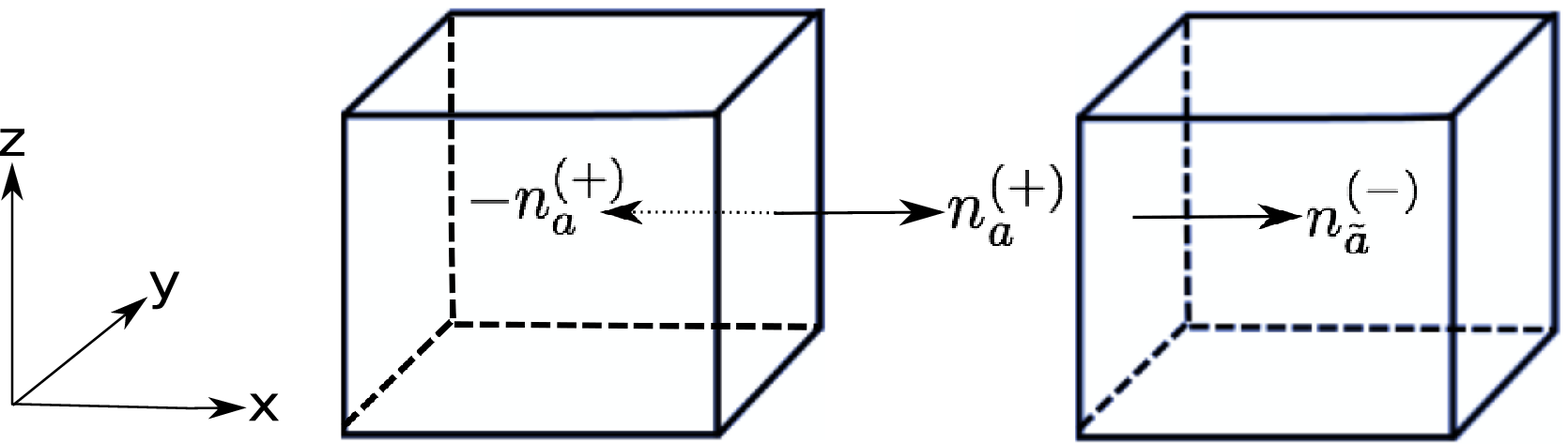}\figsubcap{b}}
 \caption{Two adjacent cubic cells, with (a) example matter content and (b) an illustration of a normal vector. The second cell is the mirror image of the first. This figure was produced using an image from Croton \emph{et al.} \cite{volk}.}
\label{fig1}
\end{center}
\end{figure}

In this section we will briefly describe how we construct a post-Newtonian cosmology. We begin by splitting the universe up into a large numbers of cells and placing them next to each other to form a periodic lattice structure. Each cell is identical to every other up to translations, rotations and reflections. At a single instance of time, the cell shape can be any regular convex polyhedron. In general, there are 11 possible ways to tesselate our universe using regular convex polyhedra\cite{vaas, poly}. The tessellation also depends on whether our universe is open, closed or flat. In Fig. \ref{fig1}(a), we have one possible example: cubic cells. The geometry inside of each cell is given by a perturbed Minkowski metric that satisfies the post-Newtonian formalism. The details of what we mean by the post-Newtonian formalism is given in Sanghai \emph{et al.}\cite{vaas}. This means that our cell size must be much less then the Hubble radius, as the boundaries of a cell must have a velocity that is much less than the speed of light. To make our model tractable, we also assume reflective symmetry across the boundary of every pair of cells. Now we can join these perturbed Minkowski, weak field patches to construct a global spacetime. 

We match these cells together using Israel junction conditions\cite{Is1} that are, in the absence of surface layers, given by
\begin{align}
 \gamma_{ij}^{(+)}  =  \gamma_{ij}^{(-)} \quad \text{and} \quad K_{ij}^{(+)} = K_{ij}^{(-)}, 
\end{align}
where $\gamma_{ij}$ is the induced metric, and $K_{ij}$ is the extrinsic curvature of the boundary, defined by
\begin{align} 
{K}_{ij}  \equiv  \frac{\partial{x^{a}}}{\partial{\xi^{i}}}\frac{\partial{x^{b}}}{\partial{\xi^{j}}} n_{a;b} \ , 
\end{align}
where $\xi^{i}$ denotes the coordinates on the boundary, and $n^{a}$ is the space-like unit vector normal to the boundary. In our model the boundary is a $2+1$ dimensional time-like hypersurface. As can be seen from Fig. \ref{fig1}(b), mirror symmetry implies that $n^{(-)}_{\tilde{a}}= -n^{(+)}_{a}$. Symmetry therefore demands that
\begin{align}  
\frac{\partial{x^{a}}}{\partial{\xi^{i}}}\frac{\partial{x^{b}}}{\partial{\xi^{j}}} n^{(+)}_{a;b} =  -\frac{\partial{x^{\tilde{a}}}}{\partial{\xi^{i}}}\frac{\partial{x^{\tilde{b}}}}{\partial{\xi^{j}}} n^{(+)}_{\tilde{a};\tilde{b}} \
\end{align}
This implies that ${K}_{ij} = - {K}_{ij}$, or, in other words, ${K}_{ij}=0$. This then allows us to evaluate the equation of motion for the boundary of a cell. As each cell is identical, this also tells how we should expect the large-scale expansion of the universe to behave\cite{vaas}.

\section{Results}
For our model we consider a late time universe filled with normal matter, where the pressure of matter is much less than the energy density. We then work out the effect of non-linear structure on the large-scale expansion of the universe. In Sanghai \emph{et al.}\cite{vaas} we have derived the equation of motion for the boundary of any regular convex polyhedra, up to post-Newtonian orders, i.e up to O($\epsilon^6$) corrections. Due to the periodicity of our model, this also tells how we expect the large-scale expansion of our universe to behave. In the specific case of cubic cells, at Newtonian order, we obtain
  \begin{align} 
\frac{(X_{,t})^2}{X^2} =   \frac{\pi G M}{3 X^3} - \frac{C}{X^2} + O(\epsilon^4) \, 
\end{align} 
 where $X(t,y,z)$ is the distance from the centre of the cell to the centre of the cell face, $M$ is the mass of the matter within a cell,  and $C$ is an integration constant that comes from the initial conditions. $X$ behaves like the scale factor and $C$ mimicks a gaussian curvature-like term, when compared to the the standard Friedmann equation for a universe filled with normal matter. 
For regularly arranged point-like masses in cubic cells, at post-Newtonian orders, we obtain
	          \begin{equation} 
 \frac{(X_{,t})^2}{X^2} =  \frac{2N}{X^3} -\frac{J}{X^4} - \frac{C}{X^2}   + O(\epsilon^6)\, 
\end{equation} 
 where $N$ and $J$ are positive constants that we have found in Sanghai et al.\cite{vaas}. At post-Newtonian orders the correction looks like a radiation term. However, this is not an actual radiation term. This term is purely due to the non-linearity of Einstein's field equations, and the inhomogeneity of our model. The size of the post-Newtonian correction depends on the size of the cell. For a cell size around the homogeneity scale \emph{i.e.} about 100 Mpc, the correction would be about $10^{-4}$ times the leading-order dust-like term. We did not need to perform any averaging to obtain our results due to the periodicity of our model.
 
So far we have only considered coordinate distances and coordinate times. However, in Sanghai \emph{et al.}\cite{vaas} we transform these quantities into coordinate independent quantities such as proper length along the edge of a cell, and proper time of an observer moving along one of the corners of these cells. The results we obtain have a similar functional form. The standard Friedmann-like behaviour that we obtain at Newtonian order is a purely emergent phenomena. This is in contrast to other models such as the Swiss cheese models that start with a FLRW background and embed Schwarzschild or Lema\^{i}tre-Tolman-Bondi patches within them\cite{ein,ltb}. However, our model has the disadvantage of only working perturbatively, whereas these models can be solved exactly. The existence of a radiation-like term has also been found previously in other models. Firstly, in the case of regularly arranged black holes in a lattice under reflective symmetric boundary conditions\cite{cgr}, and secondly in the case of the short wavelength approximation for fluctuations around a background metric\cite{wald}. 

\section{Conclusion}

We have constructed a perturbative framework that consistently tracks non-linear effects of small-scale structure on the large-scale expansion. Future developments of this work might include calculating observables in these type of models. One could also try to generalize our model by reducing the amount of symmetry that is assumed. With future large-scale surveys, such as \emph{Euclid}\cite{euclid} and \emph{SKA}\cite{ska}, we will have more data to help understand the large-scale expansion of the universe. Inhomogeneous models may then help us include non-linear gravitational effects that are usually neglected.

\section*{Acknowledgments}

I am grateful to T. Clifton for helpful comments. VAAS acknowledges support from the STFC.

\end{document}